\documentclass[twocolumn,showpacs,preprintnumbers,prl]{revtex4}
\usepackage{dcolumn}
\usepackage{bm}
\usepackage{color}
\usepackage{float}
\usepackage[caption = false]{subfig}
\usepackage[final]{graphicx}
\usepackage{amsmath}

\begin{document}

\title{Efficient Generation of Subnatural-Linewidth Biphotons by Controlled Quantum Interference}

\author{Ravikumar Chinnarasu}
\email{ravi@phys.nthu.edu.tw}
\author{Chi-Yang Liu}
\author{Yi-Feng Ding}
\author{Chuan-Yi Lee}
\author{Tsung-Hua Hsieh}
\author{Ite A. Yu}
\author{Chih-Sung Chuu}
\email{cschuu@phys.nthu.edu.tw}
\affiliation{Department of Physics, National Tsing Hua University, Hsinchu 30013, Taiwan\\
Center for Quantum Technology, Hsinchu 30013, Taiwan}

\begin{abstract}

Biphotons of narrow bandwidth and long temporal length play a crucial role in long-distance quantum communication (LDQC) and linear optical quantum computing (LOQC). However, generation of these photons usually requires atomic ensembles with high optical depth or spontaneous parametric down-conversion with sophisticated optical cavity. By manipulating the two-component biphoton wavefunction generated from a low-optical-depth (low-OD) atomic ensemble, we demonstrate biphotons with subnatural linewidth in the sub-MHz regime. The potential of shaping and manipulating the quantum wavepackets of these temporally long photons is also demonstrated and discussed. Our work has potential applications in realizing quantum repeaters and large cluster states for LDQC and LOQC, respectively. The possibility to generate and manipulate subnatural-linewidth biphotons with low OD also opens up new opportunity to miniaturize the biphoton source for implementing quantum technologies on chip-scale quantum devices. 
\end{abstract}


\maketitle

\section{I. Introduction} 

Narrowband biphotons with subnatural linewidth plays an indispensable role in photonic quantum technologies \cite{Predojevic15}. For example, the implementation of quantum repeaters \cite{Briegel98} for long-distance quantum communication or large cluster states~\cite{Browne05} for linear optical quantum computing relies on the photonic entanglement stored in quantum memories. The use of narrowband photons is thus advantageous for increasing the storage efficiency particularly in the electromagnetically-induced-transparency-based (EIT-based) quantum memories \cite{Liu01,Phillips01,Kuzmich03,Matsukevich06,Chen06,Chuu08,Zhao09a,Zhao09b,Chen13,Hsiao18,Wang19}. Moreover, the possibility to manipulate the waveform of narrowband biphotons or heralded single photons has also made possible the faithful quantum-state mapping~\cite{Cirac97}, high-efficiency quantum memory \cite{Gorshkov07,Zhang12}, the efficient loading of single photons into a cavity \cite{Liu14}, the purification of single photon~\cite{Feng17}, the measurement of ultrashort biphotons \cite{Belthangady09}, and the revival of quantum interference and entanglement \cite{Wu19}. By manipulating the phase composition of the wavefunction, narrowband single photons can also be hidden in a noisy environment~\cite{Belthangady10} and photon pairs can behave like fermions \cite{Specht09}. 

Narrowband biphotons or single photons can be realized by various mechanisms \cite{Balic05,Du08a,Srivathsan13,Zhao14,Shu16,Guo17,Kuklewicz06,Bao08,Scholz09,Wolfgramm11,Rambacha16,Wu17,Kuhn02,Keller04,McKeever04,Thompson06,Duan01,Farrera16}. However, the efficient generation of subnatural-linewidth biphotons, which is advantageous for efficient light-matter interaction at the single-photon level, typically necessitates large (relative) group delays $\tau_g \approx L/V_g$ with high-optical-depth (high-OD) atomic ensembles, where $L$ is the length of the atomic cloud and $V_g$ is the anti-Stokes group velocity, or sophisticated optical cavities with parametric down-conversion. Here, by manipulating the two-component biphoton wavefunction in a low-OD atomic ensemble, we demonstrate subnatural-linewidth biphotons with a group delay tens of times smaller than the previous works. Moreover, we achieve a biphoton linewidth in the sub-MHz regime with a limitation only imposed by the ground-state decoherence. Thanks to the possibility of controlling the quantum interference between the two-component biphotons, we also demonstrate the feasibility of shaping these biphotons and discuss their potential applications. As future quantum technologies require integrated optics architecture for improved performance and scalability, our work allows the miniaturization of subnatural-linewidth biphoton or single-photon source for chip-scale quantum devices \cite{Fortagh07,Reichel11}, where the realization of high OD is challenging.  

This paper is organized as follows. In Sec.~II we discuss the key features of the biphoton generation exploiting off-resonance coupling field in spontaneous four-wave mixing (FWM). The experimental setup is then introduced in Sec.~III. The demonstration of the controlled quantum interference and sub-MHz-linewidth biphotons are described in Sec.~IV and Sec.~V, respectively. Finally, the applications of the spontaneous FWM with off-resonance coupling field are given in Sec.~VI before we conclude our work in Sec.~VII.

\begin{figure}[t]
\centering
\includegraphics[width=1 \linewidth]{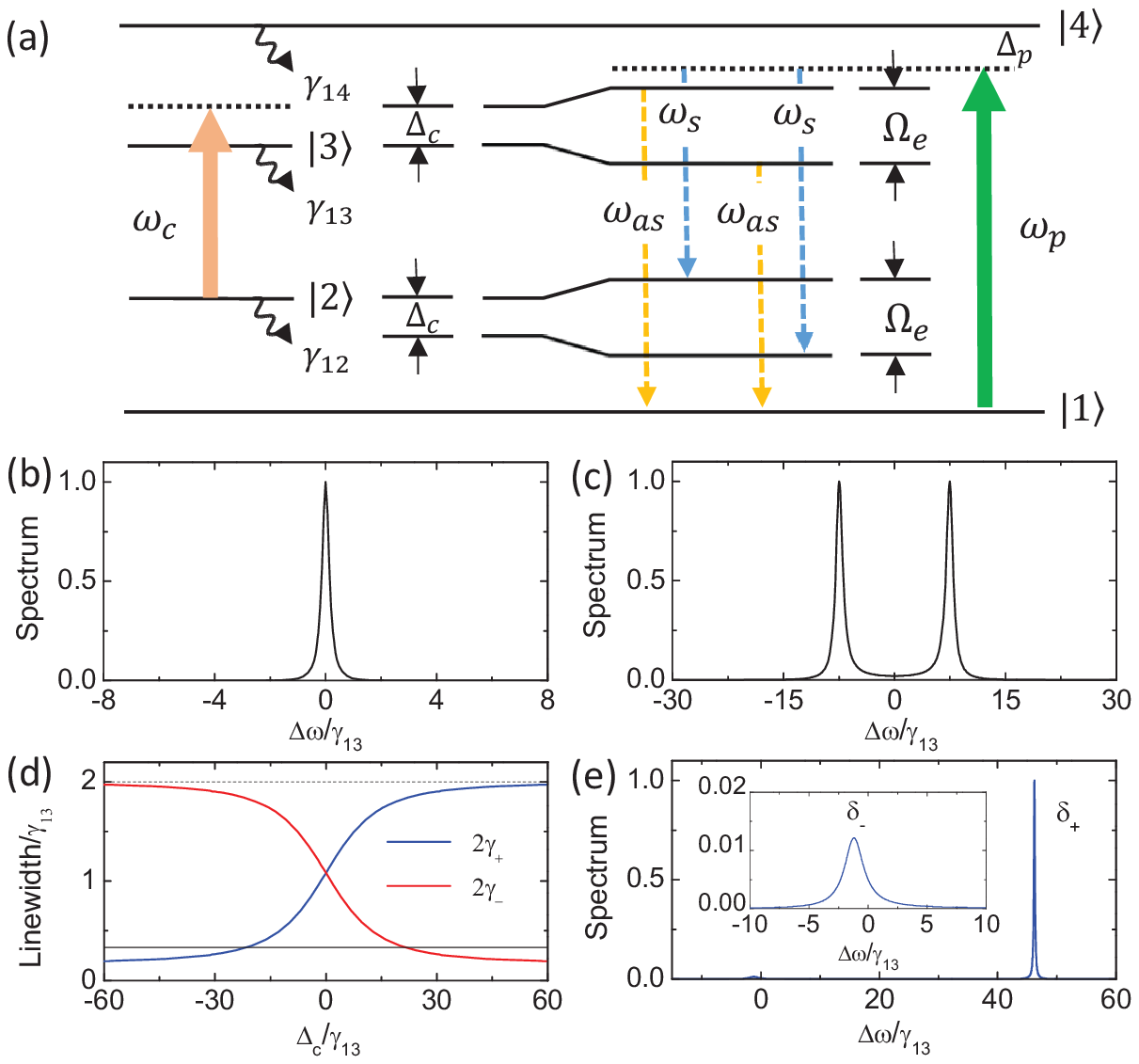}
\caption{\label{fig1} (a) Energy level diagram of the biphoton generation. If $\Delta_c = 0$, the biphotons either (b) are spectrally inseparable ($\Omega_c = 0.5 \gamma_{13}$) or (c) have linewidths limited by $\gamma_{13}+\gamma_{12}$ ($\Omega_c = 15 \gamma_{13}$). (d) If $\Delta_c \neq 0$, the biphoton linewidth is tunable by controlling $\Delta_c$. The dash and solid lines correspond to the natural linewidth and the MHz linewidth, respectively. (e) With $\Delta_c = 45 \gamma_{13}$ ($\Omega_c = 14.8 \gamma_{13}$), sub-MHz biphoton linewidth can be achieved with a dominant spectral power density.}
\end{figure}

\section{II. Biphoton generation at low optical depth} 

Fig.~\ref{fig1}(a) illustrates the energy level diagram used in our experiment, where the biphotons are generated through the spontaneous FWM \cite{Balic05,Du08a,Srivathsan13,Zhao14,Shu16,Guo17}. In contrast to the previous works, where the EIT is exploited to obtain large group delays and narrow linewidths, the group delay in our experiment is kept small compared to other time scales in the system. Consequently, the EIT does not play a significant role; the large group delay and high OD are thus not necessary for obtaining the subnatural linewidth. Furthermore, as the large-group-delay scheme necessitates high ODs to achieve high efficiency for generating narrowband biphtons, our scheme can also realize higher biphoton rate and brightness at moderate ODs. Nevertheless, the biphoton generation with small group delays typically results in either spectrally inseparable two-component biphotons or a linewidth limited by the dephasing rates of both the excited and ground states. To resolve these obstacles, we spectrally manipulate the two-component biphoton wavefunctions with an off-resonance coupling field, thus allowing us to demonstrate subnatural-linewidth biphotons in the sub-MHz regime with small group delay and low OD. 

To elaborate the importance of the off-resonance coupling field, we consider the spontaneous FWM at low OD such that the group delay of the anti-Stokes photons is smaller than other time scales in the system. The wavefunction of the biphotons generated in this regime~\cite{Wen07,Du08b} (see Appendix for more details), $\Psi (t_{as}, t_s)=\psi (\tau) {\rm exp} [-i(\omega_c+\omega_p) t_s]$, is predominantly determined by the third-order nonlinear susceptibility $\chi^{(3)}(\omega)$, where
\begin{equation}
\psi (\tau) \simeq -\frac{i \sqrt{\varpi_{as} \varpi_s}E_p E_c L}{\sqrt{8 \pi} c} \int d \omega \chi^{(3)}(\omega) e^{-i \omega \tau},
\label{Eq:psi}
\end{equation}
$\tau = t_{as} - t_s$ is the time delay between the detection of the anti-Stokes and Stokes photons, $\varpi_{as}$ ($\varpi_s$) is the center frequency of the anti-Stokes (Stokes) photons and $E_p$ ($E_c$) is the amplitude of the pump (coupling) field. For $\Omega_c \gg |\gamma_{13}-\gamma_{12}|$,
\begin{eqnarray}
&&\chi^{(3)}(\omega) \simeq \frac{-N\mu_{13}\mu_{32}\mu_{24}\mu_{41}/[4\varepsilon_0\hbar^3(\Delta_{p}+i\gamma_{14})]}{(\omega - \delta_- + i \gamma_+)(\omega - \delta_+ + i \gamma_-)}, \label{Eq:chi3} \\
&&\delta_{\pm} = \frac{1}{2} \left( \Delta_c \mp \Omega_e \right), \label{Eq:delta} \\
&&\gamma_{\pm} = \frac{( \gamma_{13} + \gamma_{12})}{2} \pm \frac{\Delta_c}{\Omega_e} \frac{(\gamma_{13} - \gamma_{12})}{2}, \label{Eq:linewidth}
\end{eqnarray}
where $\Delta_p$ ($\Delta_c$) is the detuning of the pump (coupling) field, $\Omega_e = \sqrt{|\Omega_c|^2 + \Delta_c^2}$ is the effective Rabi frequency, $\gamma_{1j}$ is the dephasing (spontaneous decay) rate of the state $\vert j\rangle$, $N$ is the atomic density and $\mu_{lm}$ is the atomic dipole moment of the $|l\rangle\leftrightarrow|m\rangle$ transition. The biphotons are thus composed of two frequency components resulting from two possible paths of FWM with red- and blue- detuned anti-Stokes photons (detuning of $\delta_+$ and $\delta_-$), respectively. If the coupling field is on-resonance, the two-component biphotons either are spectrally inseparable [Fig.~\ref{fig1}(b)] or have linewidths limited by $\gamma_{13}+\gamma_{12}$ [Fig.~\ref{fig1}(c)].

\begin{figure}[t]
\includegraphics[width=1 \linewidth]{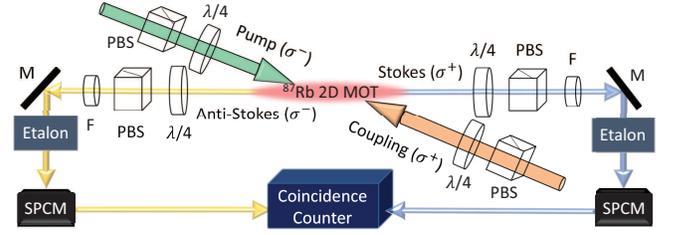}
\caption{\label{fig2} Experimental setup. $\lambda /4$: quarter-wave plates, PBS: polarizing beamsplitters, F: bandpass filters, M: mirrors, SPCM: single-photon counting modules.}
\end{figure}

If we introduce a detuning in the coupling field, the linewidths of these two components ($2\gamma_-$ and $2\gamma_+$) are both subnatural and can be tuned by controlling the coupling detuning $\Delta_c$. This can be seen in Fig.~\ref{fig1}(d), where we calculate the linewidths for $\gamma_{13} = 2\pi \times 3$~MHz, $\gamma_{12} = 0.084 \gamma_{13}$ and $\Omega_c = 14.8 \gamma_{13}$. For $\Delta_c > \Omega_c$ ($\Delta_c < -\Omega_c$), the linewidth $2\gamma_-$ ($2\gamma_+$) approaches $2 \gamma_{12} = 0.17 \gamma_{13}$ and is in the sub-MHz regime with a dominant spectral power density [Fig.~\ref{fig1}(e)]. Compared to the large-group-delay scheme, the ultra-narrow linewidth is thus achieved without the need of large group delay or high OD. In addition, the generation of the subnatural-linewidth biphotons is also more efficient at moderate ODs. For example, to achieve a biphoton linewidth of 0.3 MHz at $\gamma_{12}=0.04 \gamma_{13}$, the generation rate $R \sim |\Psi (t_{as}, t_s)|^2 t_c$ ($t_c$ is the time bin) with OD = 5 will be 35-times higher than that of the large-group-delay scheme \cite{Du08b} with OD = 100. To put it another way, the large-group-delay scheme would need an OD higher than 3000, which is very challenging if not impossible with current technology, to obtain a similar rate.

\section{III. Experimental setup} 

\begin{figure}[t]
\centering
\includegraphics[width=1 \linewidth]{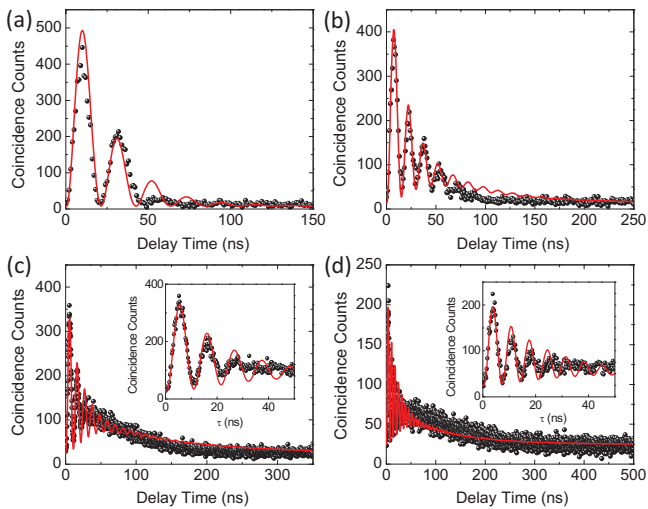}
\caption{\label{fig3} Biphoton wavepackets with (a) OD = 5, $\Delta_c = 0$ and $\Omega_c = 14.8 \gamma_{13}$, (b) OD = 5, $\Delta_c = 16.7\gamma_{13}$, and $\Omega_c = 14.8 \gamma_{13}$, (c) OD = 5, $\Delta_c = 28.3\gamma_{13}$, and $\Omega_c = 14.8 \gamma_{13}$ and (d) OD = 5, $\Delta_c = 45\gamma_{13}$, and $\Omega_c = 14.8 \gamma_{13}$. The measurement times are 300, 600, 3600, and 7200 s in (a-d), respectively. The time bins are 1 and 0.25~ns in (a-b) and (c-d), respectively.}   
\end{figure}

The schematic of our experimental setup is illustrated in Fig.~\ref{fig2}. Time-energy-entangled photons are generated using an elongated cloud of $^{87}$Rb atoms in a two-dimensional magneto-optical trap (2D MOT) with an OD of 5, where the relevant atomic levels are $\vert 1\rangle = \vert 5S_{1/2}, F=1\rangle$, $\vert 2\rangle = \vert 5S_{1/2}, F=2\rangle$, $\vert 3\rangle = \vert 5P_{1/2}, F=2\rangle$ and $\vert 4\rangle = \vert 5P_{3/2}, F=2\rangle$. More specifically, we drive the atomic ensemble with two counter-propagating fields at a tilt angle of 2.5$^{\rm o}$: the $\sigma^+$-polarized coupling field $\omega_c$ is detuned from $|2\rangle\leftrightarrow|3\rangle$ by $\Delta_c$ and the $\sigma^-$-polarized pump field $\omega_p$ is red-detuned from $|1\rangle\rightarrow|4 \rangle$ by 14$\gamma_{13}$. Through the FWM, phase-matched Stokes ($\omega_s$, $\sigma^+$-polarized) and anti-Stokes ($\omega_{as}$, $\sigma^-$-polarized) photons are spontaneously generated in the backward-wave configuration. The generated Stokes and anti-Stokes photons pass through a set of polarization and band-pass filters and polarization-maintaining single-mode fibers with a coupling efficiency of $75\%$. Among these filters, we use a Fabry-Perot etalon with 15-MHz bandwidth in the anti-Stokes channel to spectrally select the subnatural-linewidth biphotons out of the two FWM channels. The temperature-stabilized etalon has a plano-convex geometry with a transmission and FSR of 12\% and 22.9 GHz, respectively. The time-resolved coincidence counts of the biphotons are then registered by two single-photon counting modules ($60\%$ quantum efficiency) and a time-to-digital analyzer (0.25 or 1 ns time bin) for analyzing their temporal profiles. 

\section{IV. Controlled quantum interference} 

To generate the subnatural-linewidth biphotons at low OD, it is important to control the two-component biphoton wavefunction $\Psi(t_{as},t_s)$ from the two possible FWM channels. Experimentally, this is observed by the Glauber correlation function (see Appendix for more details),
\begin{eqnarray}
G^{(2)}(\tau)&=&|\Psi(t_{as},t_s)|^2 \propto [\ e^{-2\gamma_+\tau}+ e^{-2\gamma_-\tau} \nonumber \\
&&-2\cos(\Omega_e\tau)e^{-(\gamma_++\gamma_-)\tau} \ ],
\label{Eq:G2}
\end{eqnarray} 
where the two components exhibit the exponential decays with widths inversely proportional to their linewidths. The quantum interference is revealed by their beating, of which the frequency and width are determined by $\Omega_e$ and $\gamma_++\gamma_-$, respectively. By changing the coupling detuning, we can thus control the ratio, linewidths, and beating of the two components. We demonstrate such ability by replacing the narrowband etalon (15-MHz bandwidth) in the anti-Stokes channel with a broadband etalon (500-MHz bandwidth) in order to observe their quantum interference. The measured biphoton wavepacket for a resonant coupling field is shown in Fig.~\ref{fig3}(a), which exhibits a beating pattern with a period equal to the frequency difference of the corresponding anti-Stokes photons, $\delta_+ - \delta_- = \Omega_c$. With $g_{s,as}(\tau)$ and $g_{s,s}(\tau)$ [or $g_{as,as}(\tau)$] denoting the normalized cross- and auto-correlation functions, respectively, we obtain $C(\tau) = g^2_{s,as}(\tau)/ g_{s,s}(0) g_{as,as}(0) > 1$ with a peak value of $C_{max}=777\pm75$ at the maximum correlation, which violates the Cauchy-Schwarz inequality~\cite{Clauser74} and confirms the nonclassical correlation between the Stokes and anti-Stokes photons. The temporal correlation also exhibits beating pattern with period equal to the frequency difference of the corresponding anti-Stokes photons, $\delta_+ - \delta_- = \Omega_c$, thus manifesting the presence of quantum interference between two possible paths of FWM. The biphoton wavepacket can be described by the Glauber correlation function (see Appendix for more details),
\begin{equation}
G^{(2)}(\tau) \propto e^{-(\gamma_{13}+\gamma_{12})\tau}[1-\cos(\Omega_c\tau)],
\label{Eq:G2}
\end{equation}
as given by the red curve in Fig.~\ref{fig3}(a) with OD = 5, $\Omega_c = 14.8 \gamma_{13}$ and $\gamma_{12}=0.084\gamma_{13}$. As we increase the detuning of the coupling field $\Delta_c$, the frequency difference $\delta_+ - \delta_- = \Omega_e$ between the two possible anti-Stokes or Stokes fields increases; the period of the beating thus decreases. This is evident in Figs.~\ref{fig3}(b-d) for the detunings of $16.7\gamma_{13}$, $28.3\gamma_{13}$, and $45\gamma_{13}$, respectively, where $C_{max} = 126\pm13, 48\pm5$, and $20\pm3$. The observed wavepackets are in good agreement with the theory (red curves) for OD = 5, $\Omega_c=14.8\gamma_{13}$ and $\gamma_{12}=0.084\gamma_{13}$. Moreover, we observe the increase of the temporal length of biphoton wavepackets to 95, 150, and 175 ns in Figs.~\ref{fig3}(b-d), respectively--an indication that the linewidth of the biphotons with blue-detuned anti-Stokes photons narrows as we increase the detuning [Eq.~(\ref{Eq:linewidth})]. In addition, the beating area also reduces as the detuning increases, which is in consistent with the theory [Eqs.~(\ref{Eq:chi3}-\ref{Eq:linewidth})] that a mismatch of linewidths or temporal lengths exists between the biphotons generated from two possible FWM.
 
\section{V. Subnatural-linewidth biphotons} 

\begin{figure}[t]
\centering
\includegraphics[width=0.8 \linewidth]{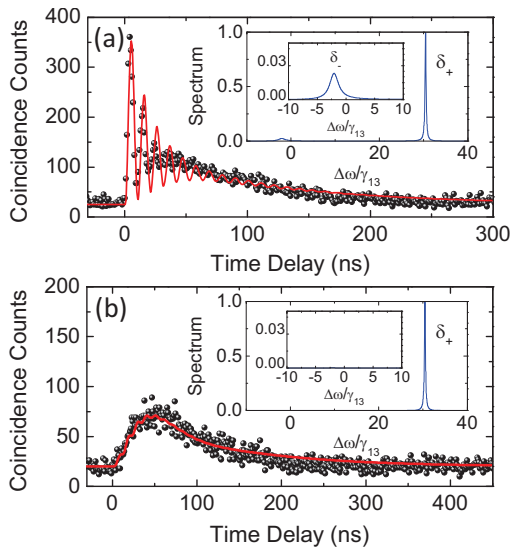}
\caption{\label{fig4} Biphoton wavepackets (a) without and (b) with the red-detuned anti-Stokes photons being discarded. The dots and curves are the experimental data (1-ns time bin and 7500-s measurement time) and theoretical fits, respectively. The corresponding spectra (arb. units) are shown in the insets.} 
\end{figure}

The observed prolongation of wavepacket's temporal length verifies the feasibility to generate narrowband biphotons using low OD. For example, with coupling detunings of $16.7\gamma_{13}$, $28.3\gamma_{13}$, and $45\gamma_{13}$ in Figs.~\ref{fig3}(b-d), respectively, subnatural linewidths of $0.42\gamma_{13}$, $0.28\gamma_{13}$, and $0.22\gamma_{13}$ can be obtained if the biphotons with red-detuned anti-Stokes photons are discarded. To demonstrate this, we spectrally select the blue-detuned anti-Stokes photons using a Fabry-Perot etalon with 15-MHz bandwidth in the anti-Stokes channel. In addition, $\Omega_c =16\gamma_{13}$ and $\Delta_c =28.3\gamma_{13}$ are chosen so that the biphotons with blue- and red-detuned anti-Stokes photons have a frequency difference large enough to be distinguished by the narrowband etalon filter. Fig.~\ref{fig4}(a) shows the biphoton wavepacket without the etalon filter at OD = 5. The quantum interference between the biphoton wavefunction from two possible FWMs is clearly observed with the beating frequency given by $\delta_+ - \delta_-$. With the insertion of the narrowband etalon filter transmitting only the blue-detuned anti-Stokes photons, the interference disappears as shown in Fig.~\ref{fig4}(b) where $C_{max} = 5\pm1$. The resulting biphotons are now single-frequency with a sub-MHz-linewidth of $0.28\gamma_{13} \simeq 859$~kHz and $\tau_g = 16$~ns as obtained by fitting the measured biphoton wavepacket (black dots) with the theory (red curve), which takes into account the measured OD (5) and ground-state decay rate (252~kHz in the presence of magnetic field inhomogeneity from the magneto-optical trap). Correcting for the quantum efficiency of each detector (60\%), the transmission of the broadband and narrowband etalon filters (45\% and 2.6\% including the fiber coupling) in the Stokes and anti-Stokes channels, respectively, the transmittance in each channel (50\%), and duty cycle (20\%), the generated paired rate is 10,368 s$^{-1}$. We note that the disappearance of the beating in Fig.~\ref{fig4}(b) is also a solid evidence of the quantum interference present with two possible FWM, for example, in Figs.~\ref{fig3}(a-d) or Fig.~\ref{fig4}(a).

\begin{figure}[t]
\centering
\includegraphics[width=0.85 \linewidth]{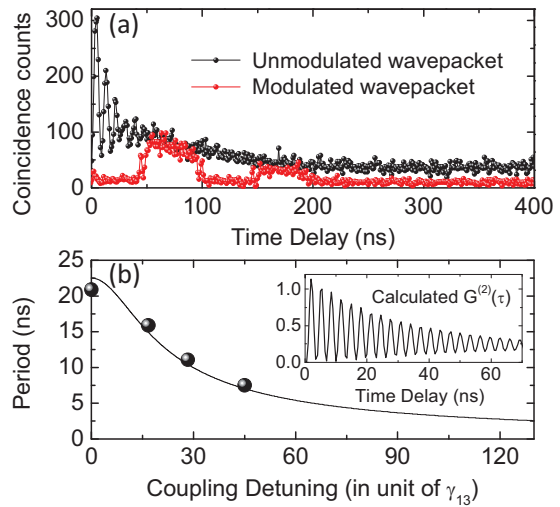}
\caption{\label{fig5} (a) Unmodulated (black dots) and modulated (red dots) single-photon wavepackets. (b) Measured periods of quantum interference as a function of the coupling detuning. The inset is the calculated wavepacket with $\Delta_c = -100\gamma_{13}$ and $\Omega_c = 30 \gamma_{13}$.}
\end{figure}
 
\section{VI. Applications} 

The controlled quantum interference demonstrated here provides an useful means for shaping the biphotons or heralded single photons generated by low-OD atomic ensembles, which were not preferred previously because of the short correlation time and the oscillatory pattern in their wavepackets. As an example in Fig.~\ref{fig5}(a) (black dots), we utilize $\Delta_c = 28.3\gamma_{13}$ and $\Omega_c =15 \gamma_{13}$ without the narrowband etalon filter to prepare single (anti-Stokes) photons heralded by the detection of Stokes photons. The single-photon wavepacket exhibits a long temporal length of 150 ns suitable for waveform or phase modulation \cite{Kolchin08}. More importantly, because of the temporal length mismatch between the wavefunctions of two possible FWM, the oscillatory pattern is limited to the front of the wavepacket and allows a large area behind for possible modulation. To modulate the wavepacket, we pass the single photons through an electro-optic modulator (20 GHz bandwidth) driven by an arbitrary function generator (80 MHz modulation frequency). The modulator is triggered by the detection signal of the Stokes photons to ensure the synchronization of the arrivals of single photons and the modulation signal. For this purpose, a 35-m-long fiber is added in the beam path of anti-Stokes photons. The modulated single-photon wavepacket with two 50-ns-long square pulses (separated by the pulse width) is shown by the red dots in Fig.~\ref{fig5}(a); the modulation would not be possible with resonant coupling field because the oscillatory pattern will occur all over the wavepacket [Fig.~\ref{fig3}(a)]. In addition to the arbitrary shaping of wavepacket, the quantum interference itself can also be utilized to realize single or entangled photons in a pulse train without the need of high-speed intensity modulators. Fig.~\ref{fig5}(b) shows the periods of quantum interference (dots) in Fig.~\ref{fig3}(a-d), which are in good agreement with the theory (curve) and tuned by controlling the coupling detuning. In the inset we also show a calculated wavepacket that exhibits a pulse train of 3-ns pulse width with $\Delta_c = -100\gamma_{13}$ and $\Omega_c = 30 \gamma_{13}$.

\section{VII. Conclusion} 

In summary, we have demonstrated biphotons with subnatural linewidth in the sub-MHz regime utilizing atomic ensemble with low OD. This is achieved by spectrally manipulating the two-component biphoton wavefunctions from two possible FWM channels. The biphoton linewidth is only limited by the ground-state decoherence in the presence of the magnetic field inhomogeneity from our magneto-optical trap. By switching off the magnetic field during the biphoton generation, it is possible to reduce the ground-state decoherence further and obtain narrower linewidth. The detuned biphotons are readily applicable to storage in the ultralow-noise room-temperature quantum memory \cite{Namazi17} and can be frequency-tuned to resonance by acousto-optic modulators (typical diffraction efficiency of 80\%) if necessary. The temporally long biphotons, without the oscillatory pattern across the wavepacket, also allows the arbitrary shaping of the photons and the generation of Bell states with subnatural linewidth \cite{Liao14} using a simpler setup.

\section{ACKNOWLEDGMENTS}
The authors thank P.K. Chen, C.W. Yang, W.L. Hung and C.H. Kuo for their experimental assistance or helpful discussion. This work was supported by the Ministry of Science and Technology, Taiwan (107-2112-M-007-004-MY3 and 107-2745-M-007-001).

\appendix

\section{Appendix: Biphoton Wavepacket}

In the perturbation theory \cite{Wen07,Du08b}, the biphoton state is given by
\begin{equation}
|\Psi\rangle=-\frac{i}{\hbar}\int_{-\infty}^{+\infty}dt\hat{H}_I|0\rangle,
\end{equation}
where the interaction Hamiltonian is
\begin{equation}
H_{I}=\frac{\varepsilon_0 A}{2} \int_{-L/2}^{L/2} dz \chi^{(3)} E_{p}^{(+)}E_{c}^{(+)}\hat{E}_{s}^{(-)}\hat{E}_{as}^{(-)}+{\rm h.c.},
\end{equation} 
$A$ is the single-mode cross-section area, in which the generated fields are collected for the correlation measurement, and $L$ is the length of the atomic ensemble. The positive-frequency part of the pump and coupling fields are described by 
\begin{eqnarray}
E^{(+)}_p(z,t)&=&E_pe^{i(k_p z-\omega_p t)},\nonumber \\
E^{(+)}_c(z,t)&=&E_ce^{i(-k_c z-\omega_c t)},
\label{Eq:Epc}
\end{eqnarray}
where $E_{p}$ and $E_{c}$ are the corresponding electric field amplitudes. $\hat{E}_{s}^{(-)}$ and $\hat{E}_{as}^{(-)}$ are the single-transverse-mode operators of the Stokes and anti-Stokes fields, 
\begin{eqnarray}
\hat{E}^{(-)}_s(z,t)=\sqrt{\frac{\hbar \varpi_s}{c \varepsilon_0 \pi A}} \int d\omega \hat{a}_s^\dagger(\omega)e^{i(k_s(\omega)z-\omega t)}, \nonumber \\
\hat{E}^{(-)}_{as}(z,t)=\sqrt{\frac{\hbar \varpi_{as}}{c \varepsilon_0 \pi A}} \int d\omega \hat{a}_{as}^\dagger(\omega)e^{i(-k_{as}(\omega)z-\omega t)},
\label{Eq:Esas}
\end{eqnarray}
Here $\hat{a}_{s}^\dagger(\omega)$ and $\hat{a}_{as}^\dagger(\omega)$ are the creation operators of the Stokes and anti-Stokes fields, respectively, which obey the commutation relations [$\hat{a}_{s}(\omega)$,$\hat{a}_{s}^\dagger(\omega')$]=[$\hat{a}_{as}(\omega)$,$\hat{a}_{as}^\dagger(\omega')$]=$\delta(\omega-\omega')$, and $\varpi_s$ ($\varpi_{as}$) is the center frequency of the Stokes (anti-Stokes) field. 

Using Eq.~(\ref{Eq:Epc}) and Eq.~(\ref{Eq:Esas}), the interaction Hamiltonian and biphoton state can be obtained as follows,
\begin{eqnarray}
\hat{H}_I&=&\frac{i\hbar L}{2\pi}\int d\omega_{as}d\omega_{s}\kappa(\omega_{as},\omega_s)\mathrm{sinc}\Big(\frac{\Delta k L}{2}\Big)\nonumber \\
&\times&\hat{a}^\dagger_{as}(\omega_{as})\hat{a}^\dagger_{s}(\omega_{s})e^{-i(\omega_c+\omega_p-\omega_{as}-\omega_s)t}
+{\rm h.c.},\\
|\Psi\rangle&=&L\int d\omega_{as} \kappa(\omega_{as},\omega_p+\omega_c-\omega_{as})\mathrm{sinc}\Big(\frac{\Delta k L}{2}\Big)\nonumber \\
&&\times\hat{a}^\dagger_{as}(\omega_{as})\hat{a}^\dagger_{s}(\omega_p+\omega_c-\omega_{as})|0\rangle,
 \end{eqnarray}
where $\Delta k=k_{as}- k_s-(k_c- k_p)$ is the phase-mismatch function, $\kappa(\omega_{as},\omega_s)=-i (\sqrt{\varpi_{as}\varpi_s}/2c)\chi^{(3)}(\omega_{as})E_pE_c$ is the nonlinear parametric coupling coefficient, and 
\begin{equation}
\chi^{(3)} (\omega)=\frac{N\mu_{13}\mu_{32}\mu_{24}\mu_{41}}{\varepsilon_0\hbar^3(\Delta_{p}+i\gamma_{14})D(\omega)}  
\label{Eq:chi3}
\end{equation}
is the third-order nonlinear susceptibility \cite{Braje04,Wen06a,Wen06b} with $D(\omega)=|\Omega_c|^2-4(\omega+i\gamma_{13})(\omega-\Delta_{c}+i\gamma_{12})$. Here $N$ is the atomic density, $\Delta_p$ and $\Delta_c$ are the detunings of the pump and coupling fields, respectively, $\gamma_{1j}$ is the dephasing (spontaneous decay) rate of the state $\vert j\rangle$, $\Omega_c= \mu_{23} E_c /\hbar$ is the Rabi frequency of the coupling field with $E_c$ being the complex amplitude of the electric field, and $\mu_{lm}$ is the atomic dipole moment associated with the transition $|l\rangle\leftrightarrow|m\rangle$.

The Glauber correlation function is defined by
\begin{eqnarray}
G^{(2)}(t_{as},t_s)&=&\langle\Psi|\hat{a}_{s}^{\dagger}(t_s)\hat{a}_{as}^{\dagger}(t_{as})\hat{a}_{as}(t_{as})\hat{a}
_{s}(t_s)|\Psi\rangle \nonumber \\
&=&|\Psi(t_{as},t_{s})|^2. 
\end{eqnarray}
Here, $\hat{a}_{s}(t_{s})$ and $\hat{a}_{as}(t_{as})$ are the annihilation operators of the Stokes and anti-Stokes fields, respectively, in the time domain. $\Psi(t_{as},t_s)$ is the biphoton wavefunction, 
\begin{equation}
\Psi(t_{as},t_s)=\psi(\tau)e^{-i(\omega_c+\omega_p)t_s},
\end{equation}
where
\begin{equation}
\psi(\tau)=\frac{L}{2\pi}\int d\omega_{as}\kappa(\omega_{as})\Phi(\omega_{as})e^{-i\omega_{as}\tau},
\end{equation}
$\tau = t_{as} - t_s$ is the time delay between the detection of the anti-Stokes and Stokes photons, and $\Phi(\omega_{as})$ is the longitudinal detuning function
\begin{equation}
\Phi(\omega_{as})=\mathrm{sinc}\Big(\frac{\Delta kL}{2}\Big)e^{i(k_{as}+k_s)L/2}.
\end{equation}
In this work, the biphoton wavefunction is dominated by ${\kappa}(\omega)$, which is associated with the third-order nonlinear susceptibility $\chi^{(3)}$, so that
\begin{equation}
\psi(\tau)\simeq -\frac{i\sqrt{\varpi_{s}\varpi_{as}} E_p E_c L}{\sqrt{8\pi}c} \int d\omega \chi^{3}(\omega)e^{-i\omega\tau}.
\end{equation}
By substituting Eq.~(\ref{Eq:chi3}) into the above equation, 
\begin{eqnarray}
\psi(\tau)\simeq C \int d\omega \frac{e^{-i\omega\tau}}{(\omega-\delta_{+}+i\gamma_{+})(\omega-\delta_{-}+i\gamma_{-})},
 \end{eqnarray}
where $C = -i\sqrt{\varpi_{s}\varpi_{as}} E_p E_c L N\mu_{13}\mu_{32}\mu_{24}\mu_{41} e^{-i\varpi_{as}\tau} /$ $\sqrt{8\pi}c \varepsilon_0\hbar^3 (\Delta_{p}+i\gamma_{14})$. The Glauber correlation function can then be evaluated using the residue theorem to be 
\begin{eqnarray}
G^{(2)}(\tau)&=&\frac{1}{2}|C|^2\big( e^{-2\gamma_+\tau}+ e^{-2\gamma_-\tau}\\ \nonumber 
&&-2\cos(\Omega_e\tau)e^{-(\gamma_++\gamma_-)\tau}\big) \Theta(\tau),
\end{eqnarray}
where $\Theta(\tau)$ is the Heaviside step function. If the coupling is on-resonance, $\Delta_c = 0$, it can be reduced to 
\begin{equation}
G^{(2)}(\tau)=\frac{1}{2}|C|^2 e^{-2(\gamma_{13}+\gamma_{12})\tau}\big[1-\cos(\Omega_c\tau)\big]\Theta(\tau).
\end{equation}


\begin{thebibliography}{99}

\bibitem{Predojevic15} \textit{Engineering the atom-photon Interaction} (Springer International Publishing 2015), A. Predojevi\'{\rm c} and M. W. Mitchell (Eds.).

\bibitem{Briegel98} H.-J. Briegel, W. D\"ur, J. I. Cirac, and P. Zoller, Phys. Rev. Lett. \textbf{81}, 5932 (1998).

\bibitem{Browne05} D. E. Browne and T. Rudolph, Phys. Rev. Lett. \textbf{95}, 010501 (2005).

\bibitem{Liu01} C. Liu, Z. Dutton, C. H. Behroozi, and L. V. Hau, Nature \textbf{409}, 490 (2001).

\bibitem{Phillips01} D. F. Phillips, A. Fleischhauer, A. Mair, R. L. Walsworth, and M. D. Lukin, Phys. Rev. Lett. \textbf{86}, 783 (2001).

\bibitem{Kuzmich03} A. Kuzmich, W. P. Bowen, A. D. Boozer, A. Boca, C. W. Chou, L.-M. Duan, and H. J. Kimble, Nature \textbf{423}, 731 (2003).

\bibitem{Matsukevich06} D. N. Matsukevich, T. Chaneli{\rm \`e}re, S. D. Jenkins, S.-Y. Lan, T. A. B. Kennedy, and A. Kuzmich, Phys. Rev. Lett. \textbf{97}, 013601 (2006).

\bibitem{Chen06} S. Chen, Y.-A. Chen, T. Strassel, Z.-S. Yuan, B. Zhao, J. Schmiedmayer, and J.-W. Pan, Phys. Rev. Lett. \textbf{97}, 173004 (2006).

\bibitem{Chuu08} C.-S. Chuu, T. Strassel, B. Zhao, M. Koch, Y.-A. Chen, S. Chen, Z.-S. Yuan, J. Schmiedmayer, and J.-W. Pan, Phys. Rev. Lett. 101, 120501 (2008).

\bibitem{Zhao09a} R. Zhao, Y. O. Dudin, S. D. Jenkins, C. J. Campbell, D. N. Matsukevich, T. A. B. Kennedy, and A. Kuzmich, Nat. Phys. \textbf{5}, 10 (2009).

\bibitem{Zhao09b} B. Zhao, Y.-A. Chen, X.-H. Bao, T. Strassel, C.-S. Chuu, X.-M. Jin, J. Schmiedmayer, Z.-S. Yuan, S. Chen, and J.-W. Pan, Nat. Phys. 5, 95 (2009).

\bibitem{Chen13} Y.-H. Chen, M.-J. Lee, I.-C. Wang, S. Du, Y.-F. Chen, Y.-C. Chen, and I. A. Yu, Phys. Rev. Lett. \textbf{110}, 083601 (2013).

\bibitem{Hsiao18} Y.-F. Hsiao, P.-J. Tsai, H.-S. Chen, S.-X. Lin, C.-C. Hung, C.-H. Lee, Y.-H. Chen, Y.-F. Chen, I. A. Yu, and Y.-C. Chen, Phys. Rev. Lett. \textbf{120}, 183602 (2018).

\bibitem{Wang19} Y. Wang, J. Li, S. Zhang, K. Su, Y. Zhou, K. Liao, S. Du, H. Yan, and S.-L. Zhu, Nat. Photon. \textbf{13}, 346 (2019).

\bibitem{Cirac97} J. I. Cirac, P. Zoller, H. J. Kimble, and H. Mabuchi, Phys. Rev. Lett. \textbf{78}, 3221 (1997).

\bibitem{Gorshkov07} A. V. Gorshkov, A. Andr$\acute{\rm e}$, M. Fleischhauer, A. S. S{\o}rensen, and M. D. Lukin, Phys. Rev. Lett. \textbf{98}, 123601 (2007).

\bibitem{Zhang12} S. Zhang, C. Liu, S. Zhou, C.-S. Chuu, M. M. T. Loy, and S. Du, Phys. Rev. Lett. \textbf{109}, 263601 (2012).

\bibitem{Liu14} C. Liu, Y. Sun, L. Zhao, S. Zhang, M. M. T. Loy, and S. Du, Phys. Rev. Lett. \textbf{113}, 133601 (2014).

\bibitem{Feng17} S.-W. Feng, C.-Y. Cheng, C.-Y. Wei, J.-H. Yang, Y.-R. Chen, Y.-W. Chuang, Y.-H. Fan, and C.-S. Chuu, Phys. Rev. Lett. \textbf{119}, 143601 (2017).

\bibitem{Belthangady09} C. Belthangady, S. Du, C.-S. Chuu, G.-Y. Yin, and S. E. Harris, Phys. Rev. A \textbf{80}, 031803(R) (2009).

\bibitem{Wu19} C.-H. Wu, C.-K. Liu, Y.-C. Chen, and C.-S. Chuu, Phys. Rev. Lett. \textbf{123}, 143601 (2019).

\bibitem{Belthangady10} C. Belthangady, C.-S. Chuu, I. A. Yu, G. Y. Yin, J. M. Kahn, and S. E. Harris, Phys. Rev. Lett. \textbf{104}, 223601 (2010).

\bibitem{Specht09} H. P. Specht, J. Bochmann, M. Mücke, B. Weber, E. Figueroa, D. L. Moehring, and G. Rempe, Nat. Photon. \textbf{3}, 469 (2009).

\bibitem{Balic05} V. Bali${\rm \acute{c}}$, D. A. Braje, P. Kolchin, G. Y. Yin, and S. E. Harris, Phys. Rev. Lett. \textbf{94}, 183601 (2005).

\bibitem{Du08a} S. Du, P. Kolchin, C. Belthangady, G. Y. Yin, and S. E. Harris, Phys. Rev. Lett. \textbf{100}, 183603 (2008).

\bibitem{Srivathsan13} B. Srivathsan, G. K. Gulati, B. Chng, G. Maslennikov, D. Matsukevich, and C. Kurtsiefer, Phys. Rev. Lett. \textbf{111}, 123602 (2013).

\bibitem{Zhao14} L. Zhao, X. Guo, C. Liu, Y. Sun, M. M. T. Loy, and S. Du, Optica \textbf{1}, 84 (2014).

\bibitem{Shu16} C. Shu, P. Chen, T. K. A. Chow, L. Zhu, Y. Xiao, M. M. T. Loy, and S. Du, Nat. Commun. \textbf{7}, 12783 (2016).

\bibitem{Guo17} X. Guo, Y. Mei, and S. Du, Optica \textbf{4}, 388 (2017).

\bibitem{Kuklewicz06} C. E. Kuklewicz, F. N. C. Wong, and J. H. Shapiro, Phys. Rev. Lett. \textbf{97}, 223601 (2006).

\bibitem{Bao08} X.-H. Bao, Y. Qian, J. Yang, H. Zhang, Z.-B. Chen, T. Yang, and J.-W. Pan, Phys. Rev. Lett. \textbf{101}, 190501 (2008).

\bibitem{Scholz09} M. Scholz, L. Koch, and O. Benson, Phys. Rev. Lett. \textbf{102}, 063603 (2009).

\bibitem{Wolfgramm11} F. Wolfgramm, Y. A. de Icaza Astiz, F. A. Beduini, A. Cer$\grave{\rm e}$, and M. W. Mitchell, Phys. Rev. Lett. \textbf{106}, 053602 (2011).

\bibitem{Rambacha16} M. Rambacha, A. Nikolova, T. J. Weinhold, and A. G. White, APL Photon. \textbf{1}, 096101 (2016).

\bibitem{Wu17} C.-H. Wu, T.-Y. Wu, Y.-C. Yeh, P.-H. Liu, C.-H. Chang, C.-K. Liu, T. Cheng, and C.-S. Chuu, Phys. Rev. A \textbf{96}, 023811 (2017).

\bibitem{Kuhn02} A. Kuhn, M. Hennrich, and G. Rempe, Phys. Rev. Lett. \textbf{89}, 067901 (2002).

\bibitem{Keller04} H. P. Keller, B. Lange, K. Hayasaka, W. Lange, and H. Walther, Nature \textbf{431}, 1075 (2004).

\bibitem{McKeever04} J. McKeever, A. Boca, A. D. Boozer, R. Miller, J. R. Buck, A. Kuzmich, and H. J. Kimble, Science \textbf{303}, 1992 (2004).

\bibitem{Thompson06} J. K. Thompson, J. Simon, H. Loh, and V. Vuleti${\rm \acute{c}}$, Science \textbf{313}, 74 (2006).

\bibitem{Duan01} L.-M. Duan, M. D. Lukin, J. I. Cirac, and P. Zoller, Nature \textbf{414}, 413 (2001).

\bibitem{Farrera16} P. Farrera, G. Heinze, B. Albrecht, M. Ho, M. Ch$\acute{\rm a}$vez, C. Teo, N. Sangouard, H. de Riedmatten, Nat. Comm. \textbf{7}, 13556 (2016).

\bibitem{Fortagh07} J. Fort$\acute{\rm a}$gh and C. Zimmermann, Rev. Mod. Phys. \textbf{79}, 235 (2007).

\bibitem{Reichel11} \textit{Atom Chips} (Wiley 2011), J. Reichel and V. Vuletic (Eds.).

\bibitem{Wen07} J. Wen, S. Du, and M. H. Rubin, Phys. Rev. A \textbf{76}, 013825 (2007).

\bibitem{Du08b} S. Du, J. Wen, and M. H. Rubin, J. Opt. Soc. Am. B \textbf{25}, C98 (2008).

\bibitem{Clauser74} J. F. Clauser, Phys. Rev. D \textbf{9}, 853 (1974).

\bibitem{Kolchin08} P. Kolchin, C. Belthangady, S. Du, G. Y. Yin, and S. E. Harris, Phys. Rev. Lett. \textbf{101}, 103601 (2008).

\bibitem{Namazi17} M. Namazi, C. Kupchak, B. Jordaan, R. Shahrokhshahi, and E. Figueroa, Phys. Rev. Appl. \textbf{8}, 034023 (2017).

\bibitem{Liao14} K. Liao, H. Yan, J. He, S. Du, Z.-M. Zhang, and S.-L. Zhu, Phys. Rev. Lett. \textbf{112}, 243602 (2014).

\bibitem{Braje04} D. A. Braje, V. Bali${\rm \acute{c}}$, S. Goda, G. Y. Yin, and S. E. Harris, Phys. Rev. Lett. \textbf{93}, 183601 (2004).

\bibitem{Wen06a} J.-M. Wen and M. H. Rubin, Phys. Rev. A \textbf{74}, 023808 (2006).

\bibitem{Wen06b} J.-M. Wen and M. H. Rubin, Phys. Rev. A \textbf{74}, 023809 (2006).


\end{thebibliography}
\end{document}